# DISCRETE SCALE RELATIVITY AND SX PHOENICIS VARIABLE STARS

Robert L. Oldershaw

12 Emily Lane

Amherst, MA 01002

USA

rloldershaw@amherst.edu

**Abstract:** Discrete Scale Relativity proposes a new symmetry principle called discrete cosmological self-similarity which relates each class of systems and phenomena on a given Scale of nature's discrete cosmological hierarchy to the equivalent class of analogue systems and phenomena on any other Scale. The new symmetry principle can be understood in terms of discrete scale invariance involving the spatial, temporal and dynamic parameters of all systems and phenomena. This new paradigm predicts a rigorous discrete self-similarity between Stellar Scale variable stars and Atomic Scale excited atoms undergoing energy-level transitions and sub-threshold oscillations. Previously, methods for demonstrating and testing the proposed symmetry principle have been applied to RR Lyrae, δ Scuti and ZZ Ceti variable stars. In the present paper we apply the same analytical methods and diagnostic tests to a new class of variable stars: SX Phoenicis variables. Double-mode pulsators are shown to provide an especially useful means of testing the uniqueness and rigor of the conceptual principles and discrete self-similar scaling of Discrete Scale Relativity.

## I. Introduction

**a. Preliminary discussion of discrete cosmological self-similarity**

The arguments presented below are based on the Self-Similar Cosmological Paradigm (SSCP)[1-6] which has been developed over a period of more than 30 years, and can be unambiguously tested via its *definitive predictions*[1,4] concerning the nature of the galactic dark matter. Briefly, the discrete self-similar paradigm focuses on nature's fundamental organizational principles and symmetries, emphasizing nature's *intrinsic hierarchical organization* of systems from the smallest observable subatomic particles to the largest observable superclusters of galaxies. The new discrete fractal paradigm also highlights the fact that *nature's global hierarchy is highly stratified*. While the observable portion of the entire hierarchy encompasses nearly 80 orders of magnitude in mass, three relatively narrow mass ranges, each extending for only about 5 orders of magnitude, account for $\geq 99\%$ of all mass observed in the cosmos. These dominant mass ranges: roughly $10^{-27}$ g to $10^{-22}$ g, $10^{28}$ g to $10^{33}$ g and $10^{38}$ g to $10^{43}$ g, are referred to as the Atomic, Stellar and Galactic Scales, respectively. The cosmological Scales constitute the discrete self-similar scaffolding of the observable portion of nature's quasi-continuous hierarchy. At present the number of Scales cannot be known, but for reasons of natural philosophy it is tentatively proposed that there are a denumerably infinite number of cosmological Scales, ordered in terms of their intrinsic ranges of space, time and mass scales. A third general principle of the new paradigm is that the *cosmological Scales are rigorously self-similar to one another*, such that for each class of fundamental particles, composite systems or physical phenomena on a given Scale there is a corresponding class of particles, systems or phenomena on all other cosmological Scales. Specific self-similar analogues from different Scales have rigorously analogous morphologies, kinematics and dynamics. When the general self-similarity among the discrete Scales is *exact*, the paradigm is referred to as Discrete Scale Relativity[5] and nature's global space-time geometry manifests a new universal symmetry principle: *discrete self-similarity or discrete scale invariance*.

Based upon decades of studying the scaling relationships among analogue systems from the Atomic, Stellar and Galactic Scales,[1-6] a close approximation to nature's discrete self-similar Scale transformation equations for the length (L), time (T) and mass (M) parameters of analogue systems on neighboring cosmological Scales $\Psi$ and $\Psi$-1, *as well as for all dimensional constants*, are as follows.

$$L_\Psi = \Lambda L_{\Psi-1} \qquad (1)$$

$$T_\Psi = \Lambda T_{\Psi-1} \qquad (2)$$

$$M_\Psi = \Lambda^D M_{\Psi-1} \qquad (3)$$

The self-similar scaling constants $\Lambda$ and D have been determined empirically and are equal to $\cong$ 5.2 x $10^{17}$ and $\cong$ 3.174, respectively.[2,3] The value of $\Lambda^D$ is 1.70 x $10^{56}$. Different cosmological Scales are designated by the discrete index $\Psi$ ($\equiv$ …, -2, -1, 0, 1, 2, …) and the Atomic, Stellar and Galactic Scales are usually assigned $\Psi = -1$, $\Psi = 0$ and $\Psi = +1$, respectively.

The fundamental self-similarity of the SSCP and the recursive character of the discrete scaling equations suggest that nature is an infinite discrete fractal, in terms of its morphology, kinematics and dynamics. The underlying principle of the paradigm is *discrete scale invariance* and the physical embodiment of that principle is the discrete self-similarity of nature's physical systems. Perhaps the single most thorough and accessible resource for exploring the SSCP and Discrete Scale Relativity is the author's website.[6]

**b. Discrete self-similarity of variable stars and excited atoms**

Discrete Scale Relativity hypothesizes that each well-defined class of systems on a given cosmological Scale $\Psi$ has a discrete self-similar class of analogue systems on any other cosmological Scale $\Psi \pm x$. Given their mass ranges, radius ranges, frequency ranges, morphologies and spherically harmonic oscillation phenomena, Discrete Scale Relativity uniquely, unambiguously and quantitatively identifies variable stars as discrete scale invariant analogues of excited atoms undergoing energy-level transitions, or oscillating with sub-threshold amplitudes at the allowed frequencies of a limited set of energy-level transitions. The latter subclass of multiple-period/low-amplitude oscillators can be interpreted as systems in sub-threshold *superposition* states.

In four previous papers the discrete self-similarity among three different classes of variable stars and their classes of analogue systems on the Atomic Scale was quantitatively demonstrated and empirically tested. Paper (I)[7] identified RR Lyrae variables as analogues of helium atoms on the basis of their narrow mass range. Given that one atomic mass unit (amu) equals approximately $1.67 \times 10^{-24}$ g, one can use Eq. (3) to determine that one stellar mass unit (SMU) equals approximately 0.145 $M_\odot$. The average mass of a RR Lyrae star ($\approx 0.6\ M_\odot$) can then be divided by 0.145 $M_\odot$/SMU to yield a value of $\approx 4$ SMU. Using the period (P) distribution for RR Lyrae variables, the $p_n \approx n^3 p_0$ relation for Rydberg atoms, and the temporal scaling of Eq. (2), one can identify the relevant principal quantum numbers (n) for RR Lyrae stars as predominantly n = 7 to n = 10. Because RR Lyrae variables are primarily fundamental radial-mode oscillators, we can assume that the most probable angular quantum numbers (l) are l = 0 or l = 1. It was further assumed that the analogue transitions are most likely to be single-level transitions. To test the foregoing conceptual and quantitative analysis of the discrete self-similarity between RR Lyrae stars and neutral helium atoms in Rydberg states undergoing single-level transitions between n = 7 and n = 10, with l = 0 or 1, a high resolution[8] period spectrum for 84 RR Lyrae stars was compared with a predicted period spectrum derived by scaling the empirical helium transition data in accordance with Eq. (2). Ten separate peak and gap structures identified in the stellar period spectrum were found to correspond quantitatively to counterpart peaks and gaps in the scaled helium period spectrum.

In Paper (II)[9] the same analysis was extended to a very large sample of over 7,600 RR Lyrae stars, and again the quantitative match between the observed and predicted period spectra supported the proposed discrete self-similarity hypothesis, albeit with some loss of resolution due to the sample size.

In Paper (III)[10] the same analytical approach was applied to high-amplitude δ Scuti variables. This class of variable stars was found to correspond to a heterogeneous class of systems with masses ranging from 10 SMU to 17 SMU. The period distribution indicated transitions in the n = 3 to n = 6 range and l values were again in the 0 to 1 range. A specific δ Scuti variable, GSC 00144–03031, was identified as a neutral carbon atom undergoing a $^{12}$C [$1s^2 2s^2 2p5p \rightarrow 4p$, (J=0), $^1S$] transition. Once again, there was an apparently unique agreement between the observed oscillation period for the variable star and the predicted period derived from the atomic data for the identified atomic analogue.

Finally, in Paper (IV)[11] the enigmatic class of ZZ Ceti variable stars was analyzed within the context of Discrete Scale Relativity. Two properties of this class of variable stars argue that it is much more heterogeneous than the RR Lyrae class or the δ Scuti stars. Firstly, the mass range of roughly 0.6 $M_\odot$ to 1.10 $M_\odot$ corresponds to atoms with masses in the 4 to 8 amu range, i.e., $^{4,6}$He, $^{6,7,8}$Li, $^{7,8}$Be and $^8$B, with the overwhelming majority expected to be analogues of $^4$He and $^7$Li.  Secondly, the broad and relatively erratic period spectrum for this class of variables has the appearance of a heterogeneous collection of discrete periods.  The fact that these white dwarf stars usually have radii well below the Stellar Scale Bohr radius of 2.76 x 10$^9$ cm strongly suggests that they correspond to ions that have lost one or more electrons.  The fact that so many ZZ Cetis are multi-mode pulsators and that the pulsations have quite low amplitudes suggests that they are equivalent to highly perturbed ions undergoing sub-threshold oscillations at allowed transition frequencies for transitions with $2 \leq n \leq 8$, $1 \leq \Delta n \leq 6$, and $1 \leq l \geq 7$.  Unique tests of Discrete Scale Relativity using ZZ Ceti stars are made very difficult by the heterogeneity of this class of variables.  However, the observed and predicted oscillation periods of a subsample of low-mass He$^+$ analogues showed good correspondence, whereas the observed oscillation periods for a high-mass ZZ Ceti subsample did not match up with the predicted period spectrum based on He$^+$ data, as would be expected.  More definitive tests of discrete cosmological self-similarity involving ZZ Ceti stars are a future research goal

### c.  SX Phoenicis Variable Stars

Fortunately, the SX Phoenicis class of variable stars are high-amplitude, low-l, oscillators that are more like the RR Lyrae stars and far less heterogeneous than the ZZ Ceti stars. They are Population II field stars with masses in a range of roughly 1.4 M$_\odot$ to 2.0 M$_\odot$ which is similar to the mass range of δ Scuti stars, but with shorter periods and higher amplitudes than is typically the case with the δ Scuti stars of Population I.  Given a reasonably accurate mass estimate for a specific SX Phoenicis star, and a knowledge of its basic oscillation properties, we can readily identify its Atomic Scale analogue and make testable predictions about specific energy-level transitions.  In the present paper, three SX Phoenicis variables: DY Pegasi, BL Camelopardalis and QU Sagittae, will be analyzed and tested using the concepts and quantitative scaling of Discrete Scale Relativity.

## II. DY Pegasi

The SX Phoenicis variable star DY Pegasi has a mass of approximately 1.5 $M_\odot$ and an effective temperature ($T_{eff}$) of approximately 7660 Kelvin.[12] The star exhibits a regular radial-mode pulsation at a frequency of 13.713 cycles per day, with a relatively large amplitude of 0.2455 magnitudes.[13] In Table 1 the basic physical characteristics of DY Peg are summarized.

**Table 1  Physical Properties of DY Peg**

| | |
|:---:|:---:|
| Mass (M) | ~ 1.5 $M_\odot$ |
| Effective Temperature ($T_{eff}$) | ~ 7660 K |
| Frequency (ν) | 13.713 $d^{-1}$ |
| Period (P) | 0.072926 d = 6300.82 sec |
| Amplitude (A) | 0.2455 mag |
| Oscillation Mode | radial |

Given the basic physical properties of DY Peg, we can use Discrete Scale Relativity (DSR) to determine the specific Atomic Scale analogue for this specific Stellar Scale system, and we can identify a very limited set of two specific atomic E-level transitions that could possibly be consistent with a rigorous analogy to the star's pulsation characteristics. The DSR analysis can then be tested quantitatively by the prediction that one, and only one, of the two-member set of possible atomic E-level transitions will have a period (p) that is related to the period (P) of DY Peg in the discrete self-similar manner required by Eq. (2), i.e., $P = \Lambda p$.

Given the star's mass of $\approx 1.5\ M_\odot$ and the DSR relation: 1 SMU = 0.145 $M_\odot$, one can determine that DY Pegasi is a 10 SMU system and that the most likely Atomic Scale analogue with 10 amu is an excited Boron atom. Given the star's period of approximately 6300.82 sec, one can determine the relevant value of n using the Stellar Scale analogue of the n versus p relation for Rydberg atoms: $P_n = n^3 P_0$, where $P_0$ for the Stellar Scale is $\Lambda p_0$ or $\approx 78$ sec. The value of n derived by this approximate method is $\approx 4.3$, and so one can specify that the relevant E-level transition is predominantly associated with n = 4. Given that the pulsation of DY Peg is primarily a fundamental radial-mode oscillation, one can safely assume that the relevant E-level transition must involve l values of 0 or 1.

The two energy-level transitions for Boron that would fit our n and l requirements are the [$1s^2\ 2s^2\ \mathbf{5s} \rightarrow 1s^2\ 2s^2\ \mathbf{4s}$] transition and the [$1s^2\ 2s^2\ \mathbf{4p} \rightarrow 1s^2\ 2s^2\ \mathbf{4s}$] transition.[13] The latter transition has a $\Delta E$ of 2776.826 cm$^{-1}$, which corresponds to an oscillation frequency ($\nu$) of 8.303 x 10$^{13}$ sec$^{-1}$ and an oscillation period (p) of 1.204 x 10$^{-14}$ sec. The question then is whether $P = \Lambda p$ for this specific E-level transition. In fact, $\Lambda p = 6263.02$ sec, which agrees with the star's P of 6300.82 sec at the 99.4% level. The [$\mathbf{5s} \rightarrow \mathbf{4s}$] transition yields a predicted P of 3386.74 sec which does not agree with 6300.82 sec. Therefore, one can conclude that the [$1s^2\ 2s^2\ \mathbf{4p} \rightarrow 1s^2\ 2s^2\ \mathbf{4s}$] transition for Boron is the unique E-level transition identified by Discrete Scale Relativity as the Atomic Scale analogue for the high-amplitude fundamental oscillation exhibited by DY Pegasi. The factor of 0.006 difference between P and $\Lambda p$ has three possible origins. Firstly, ambient Stellar Scale electromagnetic fields can shift the E-levels of systems in Rydberg states.

Secondly, the mass of DY Peg indicates that it is an analogue to $^{10}$B while the atomic data used in our calculations is based on the more common isotope $^{11}$B. Finally, and perhaps most likely, there is a small inevitable uncertainty in the value of Λ due to the fact that it is determined empirically and is therefore still a first approximation.[3] The physical properties of the proposed Atomic Scale analogue system for the DY Peg system are given in Table 2.

**Table 2   Physical Properties of the Atomic Scale Analogue of DY Pegasi**

| | |
|---|---|
| Mass | 10 amu (Boron) |
| n | ~ 4 |
| l | ~ 0 |
| ΔE | 2776.826 cm-1 |
| ν | 8.303 x 10$^{13}$ sec$^{-1}$ |
| p | 1.204 x 10$^{-14}$ sec |
| Mode | ~ radial |
| Transition | [1s$^2$ 2s$^2$ **4p** → 1s$^2$ 2s$^2$ **4s**] |

For Atomic Scale systems it is well-known that their energies and frequencies are related by Planck's law: $E = h\nu$. Discrete Scale Relativity predicts that an analogous relationship should be equally valid for the Stellar Scale analogues of Atomic Scale systems. Therefore DSR predicts that high-amplitude variable stars that constitute systems undergoing bona fide E-level transitions should obey a general law of the form: $\Delta E = H\nu$, where $H = \Lambda^{4.174} h \approx 5.86 \times 10^{47}$ erg sec. This highly important prediction could be tested if an independent method of determining $\Delta E$ for the variable stars could be identified. In that case $\Delta E/\nu$ should always be an integral multiple of H. Future research efforts should explore the possibilities of using stellar masses, effective temperatures, luminosities, spectroscopic properties, etc. to uniquely identify and quantify individual discrete E-levels for stars.

### III. The SX Phoenicis Variable of QU Sagittae

It is statistically unlikely that the discrete self-similarity between the physical characteristics and pulsation phenomena of DY Peg and Boron [4p → 4s] was a fortuitous coincidence, but it is desirable to rule out that possibility more definitively. It is proposed here that the following successful applications of the same DSR methods to two additional SX Phoenicis variables, each a double-mode pulsator, will remove any reasonable doubt about the fact that that these SX Phe stars and their Atomic Scale Boron analogues share a discrete self-similarity that is unique and quantitative. The second SX Phe star to be discussed in this paper is located in a binary star system known as QU Sagittae.[14] This SX Phe variable is a double-mode oscillator with a relatively low-amplitude of about 0.024 magnitudes and a relatively high (> 0.8) period ratio that suggests a nonradial pulsation mode for at least one of the oscillations. The primary period $P_\alpha$ is 2407.825 sec and the secondary period $P_\beta$ is 2167.206 sec. Given the $P_n = n^3 P_0$ relation used above, one may calculate an approximate value of $n \approx 3.1$ for this multi-mode oscillator. The high period ratio suggests that one should conservatively anticipate values of l in the 0 – 2 range.[14]

These stellar characteristics allow a very straightforward and quantitative test of the DSR analysis of this double-mode SX Phe star. If the analysis is unique and correct, then Boron will have two energy-level transitions associated with n ≈ 3 and 0 ≤ l ≤ 2 whose oscillation periods are related to $P_\alpha$ and $P_\beta$ by the DSR scaling for temporal periods (Eq. 2). In fact, we find that the [$1s^2\ 2s^2\ \mathbf{4s} \rightarrow 1s^2\ 2s\ \mathbf{2p^2}$] transition has a $\Delta E_\alpha$ = 7152.55 cm$^{-1}$ and a $p_\alpha$ = 4.676 x 10$^{-15}$ sec. The value of $\Lambda p_\alpha$ equals 2431.48 sec, which agrees with $P_\alpha$ at the 99% level. Likewise, we find that the [$1s^2\ 2s\ \mathbf{2p^2} \rightarrow 1s^2\ 2s^2\ \mathbf{3s}$] transition has a $\Delta E_\beta$ of 7817.35 cm$^{-1}$ and a $p_\beta$ = 4.278 x 10$^{-15}$ sec. The value of $\Lambda p_\beta$ equals 2224.71 sec, which agrees with $P_\beta$ at the 97% level.

In the case of the SX Phoenicis star in the QU Sge binary, we have used the quantitative Stellar Scale oscillation properties of the star and DSR to successfully predict two analogous Atomic Scale oscillation periods for Boron associated with transitions involving n ≈ 3 and 0 ≤ l ≤ 2. Two additional interesting properties should be noted: for this particular set of analogues the amplitudes of the stellar oscillations are quite low and the two identified E-level transitions share the [$1s^2\ 2s\ 2p^2$] level. A question for future research is whether this variable star is in a superposition of two linked but *sub-threshold* transition periods, or whether this star is simultaneously undergoing two separate but correlated transitions. The long-term evolution of the double-mode pulsation should provide clues to answering this question.

## IV. BL Camelopardalis

The final SX Phoenicis variable star to be discussed in this paper, BL Cam, is an interesting and uniquely diagnostic multi-mode SX Phe star.[15] The fundamental radial-mode oscillation has a relatively large amplitude of about 0.146 magnitudes and a period ($P_\alpha$) of 3378.24 sec, or approximately 0.0391 days. Given the $P_n \approx n^3\ P_0$ relation as a rough guide to the relevant value of n for this star, we calculate that n ≈ 4. The large amplitude and unambiguous radial-mode of this oscillation clearly point to l = 0. Therefore we can predict that Boron will have a low-l energy-level transition strongly associated with the [$1s^2\ 2s^2\ $**4s**] level, and with a scaled oscillation period of $\Lambda p_\alpha$ ≈ 3378 sec. In fact, we find that the [$1s^2\ 2s^2\ $**5s** – $1s^2\ 2s^2\ $**4s**] transition has a $p_\alpha$ of 6.513 x $10^{-15}$ sec and a $\Lambda p_\alpha$ = 3386.74 sec. The agreement between $P_\alpha$ for this SX Phoenicis star and $\Lambda p_\alpha$ for the relevant E-level transition in Boron is at the 99.8% level.

The unique and quantitative specificity of the proposed discrete self-similarity between BL Cam and a Boron atom undergoing a [5s → 4s] transition between Rydberg states can be further verified by taking advantage of the multi-mode pulsation of this star. Several of the additional pulsation periods appear to be close multiples (e.g., $2P_\alpha$) and combinations (e.g., $P_\alpha$ + $P_\beta$) which are not particularly diagnostic. However, one pulsation period ($P_\beta$) at 2727.65 sec (or about 0.03157 days) clearly appears to be an independent oscillation period that does provide a confirmatory test of the DSR analysis. Using the $P_n \approx n^3 P_0$ relation we find that this oscillation is associated with n ≈ 3. Since it is a relatively low amplitude oscillation (≈ 0.007 mag), and since the ratio of the fundamental period and this secondary period indicates that it is a *radial first overtone* pulsation,[15] we can expect l ≈ 1 for this oscillation. Therefore we can predict that Boron will have an E-level transition associated with n ≈ 3 and l ≈ 1 that has an oscillation period $p_\beta$ which scales to approximately $\Lambda p_\beta$ ≈ 2728 sec in accordance with Eq. (2). We may further expect, on the basis of finding that the two modes of the double-mode pulsator of QU Sge appeared to be linked transitions (i.e., they shared a common E-level), that the two oscillation periods of BL Cam are also *linked*. In fact, our highly unique 4-part prediction is vindicated in a convincing manner. The [$1s^2\ 2s^2\ $**4s** → $1s^2\ 2s^2\ $**3p**] transition for Boron is: (1) associated with n ≈ 3, (2) involves a change of l from l = 0 to l = 1, (3) is linked to the fundamental oscillation period through the common 4s E-level, and (4) has a $\Lambda p_\beta$ of 2719.11 sec, which agrees with the predicted period of 2727.65 sec at the 99.7% level.

This empirical support for the theoretical hypothesis of discrete cosmological self-similarity between BL Cam and an excited Boron atom undergoing E-level transitions in the $3 \leq n \leq 5$ range, with $0 \leq l \leq 1$, certainly appears to be quite definitive. Coincidence is not a viable explanation for these successful results, and the 4-part specificity of the empirical testing would appear to minimize the possibility of subjectivity or arbitrariness in the analysis.

## V.  Conclusions

Given the masses and period data for three SX Phoenicis variables, we have been able to make predictions about specific energy-level transitions of a specific type of excited atom (Boron). Technically these predictions are diagnostic *retrodictions* since the Atomic Scale data was published before the tests were conducted. For each of the five indentified pulsation periods, the Stellar Scale oscillation period was approximately $\Lambda$ times larger than the uniquely specified Atomic Scale oscillation period, with an average agreement at the 99% level. Two of the SX Phe stars analyzed here were double-mode pulsators that allowed additional degrees of specificity in testing the discrete cosmological self-similarity between SX Phe stars and excited Boron atoms oscillating at published E-level transition frequencies. The probability is exceedingly small that five multi-part tests, involving mass, n, l, and frequency values, could succeed in this manner by chance or even by the use of subjective methodology. Therefore, a high degree of discrete self-similarity between these stellar/atomic analogues from neighboring Scales of nature's discrete cosmological hierarchy appears to have been verified quantitatively.

Some additional comments and questions for future research are as follows.

(1) Clearly stellar mass is, or should be, the definitive parameter in classifying variable stars, as is the case with atoms. Although periods, pulsation modes, temperatures and spectroscopic properties are useful diagnostic characteristics, variable stars should be ordered and classified primarily by their masses in rigorous analogy to the Atomic Scale classification of atomic elements.

(2) It will be of interest to explore the apparent coupling of double-mode pulsations, wherein a shared E-level is identified. This phenomenon has been documented for QU Sge, BL Cam and the δ Scuti star GSC 00144-03031.[10] Is this a common characteristic of all double-mode Stellar Scale variables? What physical mechanism could explain this coupling of simultaneous oscillation phenomena? Is an analogous Atomic Scale phenomena known?

(3) Are the low-amplitude oscillations of multi-mode variable stars *sub-threshold* oscillations that would be virtually unobservable in Atomic Scale systems, since they would be below the energy threshold required for emission of quanta?

(4) Planck discovered the famous relation $E = h\nu$ for atoms, and Discrete Scale Relativity predicts an equivalent relation $E = H\nu$ for the high-amplitude variable stars. We can measure stellar $\nu$ values quite accurately and we can predict that $H = (\Lambda^{D+1} h)$ or about $5.86 \times 10^{47}$ erg sec in accordance with the discrete scale invariance of DSR.[5] If there were a way to define the Stellar Scale E-levels, then the proposed $E = H\nu$ relation could be empirically tested.

Taken together the evidence for discrete self-similarity involving RR Lyrae stars, δ Scuti stars, ZZ Ceti stars, SX Phoenicis stars, and their respective Atomic Scale analogues is very strong and cannot be due to coincidence or subjective methods. Discrete scale invariance allows one to investigate equivalent systems that differ in lengths and temporal periods by a scale factor of $5.2 \times 10^{17}$, and that differ in masses and energies by a scale factor of $1.70 \times 10^{56}$. If the reality of Discrete Scale Relativity were established and accepted, then the resulting paradigmatic advance would create a very rich potential for new breakthroughs in our understanding of stellar and atomic systems.